\documentclass{ifacconf}

\usepackage{graphicx}      
\usepackage{natbib}        
\usepackage{amssymb,amsmath}
\newtheorem{def1}{Definition}
\begin{document}
\begin{frontmatter}

\title{Towards a Personalisation Framework for Cyber-Physical-Social System (CPSS)}

\author[First,Second]{Bereket Abera Yilma} 
\author[First]{Yannick Naudet} 
\author[Second]{Herv\'{e} Panetto}

\address[First]{Luxembourg Institute of Science and Technology (LIST), 5, Avenue des Hauts-Fourneaux, L-4362, Esch-sur-Alzette, Luxembourg,\\
(e-mail: (bereket.yilma, yannick.naudet)@list.lu)}
\address[Second]{Université de Lorraine, CNRS, CRAN, F-54000 Nancy, France (e-mail: herve.panetto@univ-lorraine.fr)}

\begin{abstract}  
A Cyber-Physical-Social System (CPSS) is an emerging paradigm often understood as a physical and virtual space of interaction which is cohabited by humans and sensor-enabled smart devices. In such settings, human interaction behaviour is often different from person to person and is guided by complex environmental and natural factors that are not yet fully explored. Thus, ensuring a seamless human-machine interaction in CPSS calls for efficient means of handling human dynamics and bringing interaction experience to a personal level. To this end in this paper, we propose a personalisation framework to support the design of CPSS in recognising and addressing human/social aspects.
\end{abstract}


\end{frontmatter}

\section{Introduction}
The concept of Cyber-Physical-Social System (CPSS) has gained an increasing attention over the past few years.
This is often attributed to the mass integration of smart devices in various aspects of daily life, \textit{\cite{yilma2020new}}. The CPSS paradigm mainly promotes the integration of human/social aspects in the long existing notion of Cyber-Physical System(CPS), \textit{\cite{yilma2018introduction,zeng2020survey}}. This growing interest to incorporate human/social aspects in CPS has unlocked a number of research challenges. Especially since the so called smart devices are populating almost every aspects of modern life, the need to ensure a seamless interaction while respecting important human values is a key research challenge that remains open. Previously in the work of, \textit{\cite{yilma2018introduction}} personalisation was proposed as one viable solution to address this challenge. The proposal was based on the premises that one of the main contributors to the complexity of CPSS environments originates from human dynamics. This is because human actions and behaviours are guided by several environmental and personal factors which are difficult to predict and manage compared to machines. Thus, personalisation was suggested as a means to manage (\textit{i.e.} to better understand and predict) human aspects in CPSS while keeping individual's freedom to operate. However, taking into account its complexity, the problem of personalisation in CPSS is yet to be addressed. This is partially due to the lack of uniform underlining principles to the design of CPSS environments. Despite previous efforts to lay systemic ground to the foundation of CPSS in \textit{\cite{yilma2019meta,yilma2020new}} the formalisation needs to further mature in order to characterise the emergence of complex CPSS environments.

Hence, in this paper we set out to address these challenges. The contribution of this work is two fold. The first contribution is to extend the formalisation of CPSS in an effort to characterise complex structures of emerging CPSS environments. This was done by linking the existing concepts of CPSS with System-of-Systems(SoS) principles and through designing an extended meta-model from \textit{\cite{yilma2020new} and \cite{doi:10.1080/17517575.2018.1536807}}. Ultimately the second contribution is the proposal of a personalisation framework  which formalises the problem of personalisation in CPSS context based on the established concepts and the meta-model. The framework is designed to be used in CPSS environments to help ensure a seamless human-machine interaction experience. The rest of this paper is organised as follows; Section 2 presents a brief background on CPSS and the extended concepts  followed by the new meta-model. Section 3 covers the personalisation framework elaborated by a case-study on Cobotic systems for a smart workshop setting. Finally Section 4 presents a concluding discussion and highlights feasible future research directions.  

\section{Cyber-Physical-Social System(CPSS)}\label{sec:CPSS}
A systemic formalisation to the concept of CPSS was previously proposed in \textit{\cite{yilma2020new}}. The formalisation presents a domain independent definition of CPSS grounding on the theory of systems followed by a meta-model that shows the main components and relationships leading to the emergence of CPSS as a system. In this section we extend the previously proposed concepts of the CPSS paradigm to characterise complex CPSS environments that emerge as System of Systems(SoS). We first recall definitions of CPSS and SoS. Then we propose an extended meta-model elaborating the emergence of CPSS as a SoS.
\begin{def1}\label{def:CPSS}
\textbf{Cyber-Physical-Social System (CPSS)}: ``is a system comprising cyber, physical and social components, which exists or emerges through the interactions between those components. A CPSS comprises at least one physical component responsible for sensing and actuation, one cyber component for computations and one social component for actuating social functions." \textit{\cite{yilma2020new}}
\end{def1}
From a systems engineering perspective, the notion of SoS was best described as an emergent system from at least 2 \textit{loosely coupled systems}  that are collaborating;\textit{\cite{Morel2007SystemOE}}. The earliest and most accepted definition of SOS is the one by \textit{\cite{Maier1996}} defined  as follows:
\begin{def1}\label{def:SoS}
 ``A \textbf{System-of-Systems (SoS)} is an assemblage of components which individually may be regarded as systems having \textit{Operational} and \textit{Managerial Independence} \cite{Maier1996}.
\end{def1}

 \begin{figure*}[!t]
  \centering
  \includegraphics[width= 0.6\textwidth]{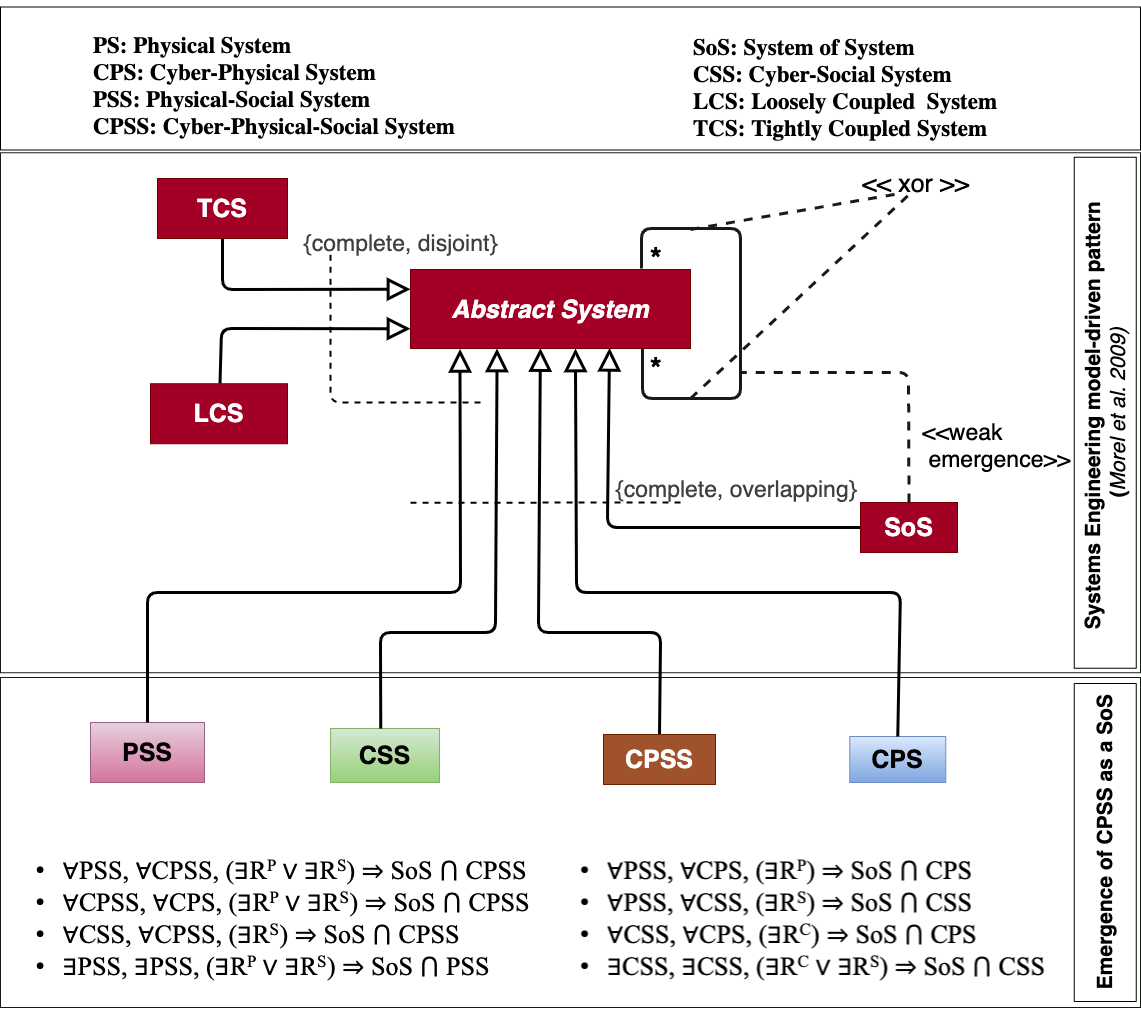}
  \caption{CPSS Meta-model}\label{fig:meta-Model}
\end{figure*}
In literature typical examples of CPSS are the so-called \textit{Smart spaces} such as \textit{smart manufacturing systems, smart homes, smart cities, etc.}  Inherently the emergence of these spaces as a CPSS is the result of the interaction of the three fundamental components (Cyber, Physical and Social) of different independent systems and humans. Here each interacting entity being independent system, has operational and managerial independence. This allows us to frame newly formed CPSS as a System of Systems(SoS) (\textit{definition \ref{def:SoS}}).  Framing CPSS as a SoS and aligning it with the theory of systems fundamentally helps to reduce the complexity of such spaces. \textit{i.e.} it helps to clearly visualise the component systems, identify their individual objectives, relationships, inter-dependencies and  determine complementary and conflicting objectives. The complexity of SoS often depends on the nature of relations between its component systems, their individual behaviour, objectives and functionalities \cite{Maier1996}. As a system a CPSS possesses key systemic properties (\textit{i.e.} components, objective, relations, behaviour, structure, interface, environment and functions). We refer the reader to the work of \textit{\cite{yilma2020new}} for a detailed explanation of the components and types of relationships leading to the emergence of CPSS. 

It is however worth recalling some of the key CPSS concepts introduced in \textit{\cite{yilma2020new}} as we set out to extend the CPSS paradigm. $\cal R$  = $\{R^{C}$, $R^{P}$, $R^{S}$, $R^{CP}$, $R^{PS}$, $R^{CS}$, and  $R^{CPS}\}$ represents the seven types of relations among components(Cyber, Physical and Social) leading to the emergence of different kinds of systems. The concept of Physical-Social System (PSS) was introduced as an emergent system from physical and social components as a result of Physical-Social relation ($R^{PS}$). An example of PSS is a human system. The rational behind the concept of PSS is to study and investigate the intangible social components (emotional, cognitive and behavioral aspects) which we eventually want to mimic in machines. 

A CPSS can take two different forms emerging as an independent system and as a SoS. The first characterises a next generation of CPS devices with an added social component enabling them to actuate socially(\textit{i.e.} detect, reason and adapt to human's emotional cognitive and behavioral responses). Whereas, the latter refers to an interaction space for humans and  smart devices. This formalism entails that humans interacting with socially constrained CPS devices form a SoS but not a true CPSS. Despite most works in literature refer to such SoS as a CPSS, social aspects need to be realised in machines for a true CPSS to emerge as a SOS. Thus, this distinction sets an evolution hierarchy for smart systems to become true CPSS. 

In order to visualise the emergence of CPSS as a SoS and also other types of SoSs formed as a result of the interactions between component systems, we present an extended meta-model using UML 2.0 notation in figure \ref{fig:meta-Model}.
  As it can be seen on the meta-model the top part illustrates concepts adopted from the work of  \textit{\cite{Morel2007SystemOE}} showing the formation of SoS as a weak emergence from the interactions between independent systems, that can be either  Tightly Coupled System(TCS) or Loosely Coupled System(LCS). The interaction link on abstract system refers to any of the relations in $\cal R$. The bottom part shows the emergence of CPSS as a SoS and also other kinds of SoSs formed in CPSS context. The axioms at the bottom illustrate the main kinds of SoSs that can be formed as a result of interactions between the independent systems.
  
  Fundamentally the postulate here is that a true CPSS is formed as a SoS when there is a social relation $R^{S}$ between a single system CPSS \textit{e.g. Cobot(Collaborative robot)} and a PSS \textit{e.g. human}. Here, having a physical relation  $R^{P}$ instead of social $R^{S}$ can form a SoS. However, it does not necessarily entail the formed SoS is a CPSS which essentially requires a social relation $R^{S}$ where the single CPSS \textit{e.g. Cobot} is able to detect, reason and adapt to social interaction responses of the human. Furthermore, CPSS can also emerge as a SoS whenever a CPS or a CSS initiate  a social relation with a single system CPSS. The first 3 axioms on Fig. \ref{fig:meta-Model} represent the basic ways a CPSS can be formed as a SoS. The rest of the axioms describe other kinds of SoSs that can be formed in a CPSS context.

In a nutshell the CPSS paradigm ultimately aims at creating smart environments where the current socially constrained CPS devices gradually evolve to understand, reason and adapt to social interaction responses of a human, thereby ensuring a seamless interaction. Doing so however requires first identifying the key social components in human-to-human interaction and mimicking those components in CPS devices. In a human-to-human interaction, it is obvious that the quality of the interaction is subject to how well the individuals know each other. (\textit{i.e.} if one knows the other person's preferences, behaviour, likes and dislikes it is more convenient to respond appropriately in a social context).  The same is true in a human-machine interaction. Having a social component by itself signifies the ability to actuate socially. However, for a seamless interaction one needs to know the interacting individual at a personal level. This is because each person is unique and his/her actions and behaviours are guided by individual skills, knowledge, preferences, interests, culture and beliefs. Hence, in the quest towards a true \textit{CPSS} the need to ensure a seamless social interaction positions the concept of  \textit{personalisation} or adaptation of the systems to human presence, at the heart of the problem. 

The gradual introduction of  personalisation and adaptation of systems in such settings poses a number of opportunities for both personalised service consumer and the CPSS. In particular it empowers smart devices by mimicking social components so  that they can have different levels of social actuation capability paving the way towards a true CPSS. Based on these premises, in the next section we present our proposal for a personalisation framework in a CPSS context. The framework is presented to serve as a basis for designing personalised and adaptable CPSS environments.
\begin{figure*}[!t]
  \centering
  \includegraphics[width= 0.6\textwidth]{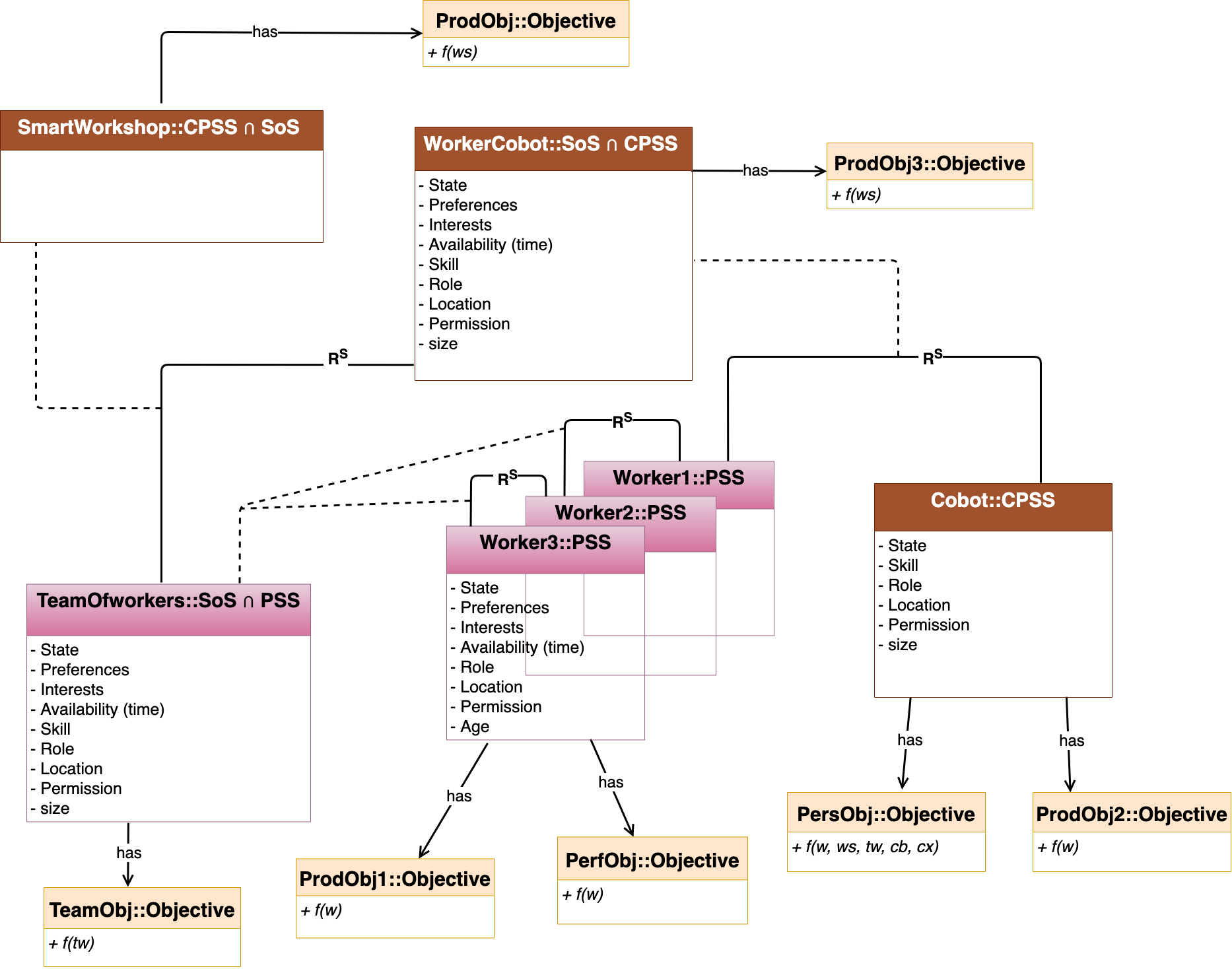}
  \caption{Conceptual model of a Smart factory based on the CPSS meta-model}\label{fig:fw}
\end{figure*}
\section{Personalisation in CPSS}\label{sec:personalisation_CPSS}
According to the discussion presented in section \ref{sec:CPSS} smart systems often seen as CPSSs are SoSs formed as a result of the interactions between independent systems. In such environments people evolve with other people and different sensor enabled devices. In personalising and making such environments adaptable to a particular person one should also consider the objectives of the co-existing entities and the global objective of the smart environment (\textit{i.e.} CPSS) \textit{\cite{yilma2018introduction,naudet2018personalisation}}. This essentially means that the personalisation should make the best possible compromise between the co-existing objectives and respect environmental constraints.  In oder to do so, one needs to first identify the main component systems that have a direct or indirect influence on the user of the personalisation service and vise versa. This leads to formalise the problem of personalisation in CPSS as a function of the main systems (\textit{i.e.} the user $u$ of personalisation service, the CPSS in which the user evolves in $cpss$, the crowd of other people in the CPSS $cr$, the application device that implements the personalisation service $d$ and the global context $cx$) written as: 
\begin{equation}\label{eqn:PuPSS}
Perso_u^{(CPSS)} = f(u,cpss,cr,d,cx)
\end{equation}
Here, the Context $cx$  refers to the set of all other elements (component systems) of the CPSS $\{x_{1}, x_{2}, ... x_{n} \}$ that have no direct or indirect influence on the user/personalisation. When any of the component systems in $cx$ has an impact on the user/personalisation it will be taken as part of the formalisation $f$ as $f(u,cpss,cr,d,x_{i}, cx)$ ; $\forall$ $x_{i}$ $\in$ $cx$.
For a more elaborated discussion in the next subsection we present a case-study of Cobotic system in a \textit{smart workshop} setting.

\subsection{Personalisation in Cobotics}

Together with advances in Industry 4.0 the use of Collaborative robots (Cobots) has become an emerging trend in various sectors. For instance in the case of Smart manufacturing systems, factories are often organised as job shops. In the production line we have  engineers, operators and maintenance technicians that are skilled and able to perform tasks on different machines. In this settings Cobots are often introduced at job shops to collaborate with the workers in order to improve efficiency. However, Cobots are often programmed to only execute predefined tasks. Hence, they are not able to adapt to changing needs of human workers. This can potentially degrades collaboration quality and could also compromise safety of human workers. By introducing  personalisation here we primarily aims at  enabling cobots to learn complex human interaction responses. Thus, they can gradually adapt to changing states respecting important human values and needs to become better companions. 

Adapting the global formalisation of personalisation in CPSS (equation \ref{eqn:PuPSS}), the problem of personalisation in Cobotics can be formalised as a function of the main systems (\textit{i.e.} the user of personalisation service translates to the worker  $w$, the CPSS which translated to the smart workshop $ws$, the crowd of other people in the factory translates to a team of workers $tw$, the device implementing the personalisation which translates to the Cobot $cb$ and the context elements $cx$) written as :
\begin{equation}\label{eqn:PuCBx}
Perso_u^{(Cob)} = f(w,ws,tw,cb,cx)
\end{equation}
In figure \ref{fig:fw} we present a conceptual model 
for the scenario of smart workshop based on the meta model presented in section \ref{sec:CPSS}.
 
 As it is depicted on the figure the class \textit{Cobot} is instantiated as a subtype of \textit{CPSS} provided a personalisation objective (\textit{PersObj})  and a production objective(\textit{ProdObj2}). The class \textit{Worker} is an instance of \textit{PSS}. Whereas the class \textit{WorkerCobot} represents a CPSS which is a SoS that emerges as a result of the relations $R^{P}$ and $R^{S}$ between a worker(PSS) and a Cobot(CPSS) according to axiom 1 on figure \ref{fig:meta-Model}.  
 The class \textit{TeamOfworkers} is another emergent SoS formed as a result of $R^{P}$ and $R^{S}$ relations among two or more workers. The class   \textit{SmartWorkshop} is thus, a CPSS which is a SoS formed from \textit{TeamOfworkers} and \textit{WorkerCobot} relations.  As an independent system each of these systems can have one or more objectives serving the global purpose of the smart workshop as well as personal ones.
 
In this particular scenario of a smart workshop personalisation is one objective which is implemented by the Cobot interacting with a worker. This essentially means enabling the Cobot to understand and reason dynamic human interaction responses and adapt to changing needs accordingly. In doing so the Cobot should also respect the objectives of the co-existing entities and the environment. 

Implementing this however is not a trivial task as it requires relaxing the control rules and training cobots to derive efficient representations of the humans state from high-dimensional sensory inputs, and use these to generalize past experience to new situations. Such kinds of challenging tasks are remarkably solved by humans and other animals through a harmonious combination of reinforcement learning(RL) and hierarchical sensory processing systems, \textit{\cite{serre2005object, fukushima1982neocognitron}}. This in particular has inspired the development of several RL algorithms over the years, \textit{\cite{nguyen2020deep}} used for training agents to perform complicated tasks. However, their application  was limited to domains in which useful features can be handcrafted, or to domains with fully observed, low-dimensional state spaces. Recently a novel artificial agent called deep Q-network (DQN) was proposed in the work of \textit{\cite{mnih2017methods}}. DQN can learn successful policies directly from high-dimensional sensory inputs using end-to-end reinforcement learning. DQN has been tested over various complicated tasks and was able to surpass the performance of all
previous algorithms \textit{\cite{silver2016mastering, silver2017mastering}}. It has also enabled the creation of ``AlphaGO";which is to date considered as one of the greatest breakthroughs in artificial intelligence that was able to beat the world's  most diligent and deeply intelligent human brains, \textit{\cite{chen2016evolution}}. This and other recent successes such as ``ÀlphaStar", \textit{\cite{10.1145/3319619.3321894}}  demonstrate the potential of RL to build intelligent agents by giving them the freedom to learn by exploring their environment and  make decisions to take actions which maximises a long term reward.

We believe that RL can be beneficial to the task of personalisation in CPSS as it allows agents to learn by exploring their environment unlike supervised methods which require collecting huge amount of labeled data and harder to train with continuous action space. Taking this inspiration we reformulate the task of personalisation in Cobotics as an RL task by extending the formalisation in equation \ref{eqn:PuCBx}.

\begin{figure}[!h]
  \centering
  \includegraphics[width= 0.4\textwidth]{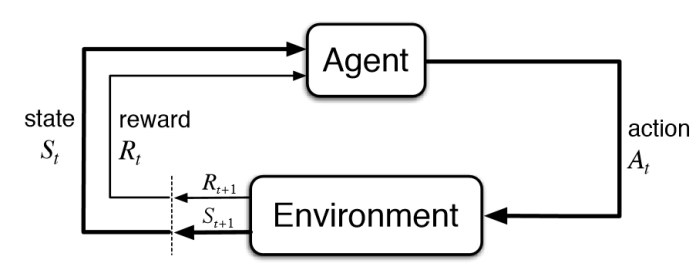}
  \caption{Reinforcement Learning(\cite{sutton1998introduction})} \label{fig:RL}
\end{figure}

\noindent
In a classical RL, agents interact with their environment through a sequence of observations, actions and rewards \cite{watkins1992q}. At a given time an agent takes observation (\textit{i.e.}information about the state of the environment) and takes an action that will maximise a long term reward. The agent then observes the consequence of the action on the state of the environment and the associated reward. It then continues to make decisions about which actions to take in a fashion that maximizes the cumulative future
reward. This is done by learning action value function,
\begin{equation}
Q^{*} (s,a) = \max_\pi \mathbb{E}\left [\sum_{t\geqslant 0}^{} \gamma^{t} r_{t}|s_{0}= s, a_{0} = a, \pi  \right ]
\end{equation}
which is the maximum sum of rewards $r_{t}$ discounted by $\gamma$ at each time step $t$, achievable by a policy $\pi=p(a\mid s)$, after making an observation of $(s)$ and taking an action $(a)$. This means that RL agents operate based on a policy $\pi$ to approximate $Q$-values(state-action pairs) that maximise a future reward.  Figure \ref{fig:RL} illustrates the schematics of the different components in classical RL. (We refer the reader to the work of \textit{\cite{watkins1992q, sutton1998introduction}} for the details on Q-learning and RL).

\begin{figure}[!h]
  \centering
  \includegraphics[width= 0.4\textwidth]{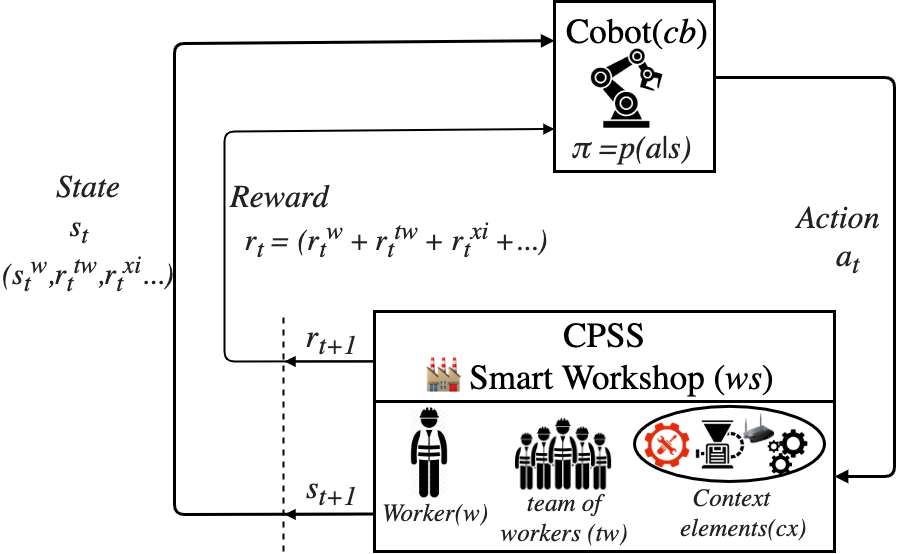}
  \caption{Personalisation in Cobotics as an RL task}\label{fig:perso_RL}
\end{figure}

Adopting this to the context of Cobotics, the Cobot corresponds to the agent which operates based on a policy $\pi$ and the environment corresponds to the smart workshop which is a CPSS containing a worker (target user of personalisation), the cobot itself, the team of workers, other context elements (\textit{i.e.} devices and objects). The state of the environment $s_t$ at any time step $t$ is a combination of the states of the main entities in the workshop (\textit{i.e.} state of the worker $s_{t}^{w}$, state of the team of workers $s_{t}^{tw}$, and state of any context element that has an impact on the worker $s_{t}^{x_i}$). Similarly the action taken by the cobot $a_t$ can be one or a combination of other actions according to the states of the respective entities depending on the scenario. The reward $r_t$ the cobot receives for taking an action $a_t$ is the total sum of the rewards deemed appropriate for the corresponding states of the main entities ($r_t$ =  $r_{t}^{w}$ +  $r_{t}^{tw}$ +  $r_{t}^{x_i}$ + ...). In RL reward values play a crucial role in guiding the exploratory behaviour of the agent (\textit{i.e.} the Cobot in our case). Since the main objective of personalisation  here is enabling the Cobot to make informed decisions and take actions adapting to needs of the worker, $r_{t}^{w}$ should be prioritised. Doing so, the cobot should not cause significant harm on the functioning of the other entities. This will be regulated by the rewards associated with the co-existing entities ($r_{t}^{tw}$, $r_{t}^{x_i}$, etc.) . Figure \ref{fig:perso_RL} illustrates the problem of personalistion in cobotics as an RL task.

In classical RL at each step the approximation of the optimal Q-value function $Q^{*}$ will be refined by enforcing the \textit{``Bellman equation"} \textit{\cite{watkins1992q}} given by:
\begin{equation}
Q^{*} (s,a) =  \mathbb{E}_{s^{'}\sim\varepsilon } \left [ r + \gamma\max_{a^{'}} Q^{*} (s^{'},a^{'})|s,a \right ],
\end{equation}
which states that given any state-action pair $s$ and $a$ the  maximum cumulative reward achieved is the sum of the reward for that pair $r$ plus the value of the next state we end up with $s^{'}$. The value at state $s^{'}$ is going to be the maximum over actions $a^{'}$ at $Q^{*} (s^{'},a^{'})$. Thus the optimal policy $\pi^{*}$ corresponds to taking the best action in any state as specified by $Q^{*}$. 
In this iterative process the Bellman equation is used as a value iteration algorithm which iteratively refines $Q^{*}$:
\begin{equation}
Q_{i+1} (s,a) =  \mathbb{E}\left [ r + \gamma\max_{a^{'}} Q_{i} (s^{'},a^{'})|s,a \right ],
\end{equation}

$Q_{i}$ converges to $Q^{*}$ as $i$ approaches to infinity. 

For the problem of personalisation in Cobotics we are interested in finding an optimal policy on which the Cobot operates on in order to take the best possible action given the state of the workshop (\textit{i.e.} $s_{t}^{w}$, $s_{t}^{tw}$,$s_{t}^{x_i}$). Since  workers in such settings experience mental as well as physical workloads they often produces a subjective experience and respond differently depending on individual skills, characters, preferences, etc. In particular our main interest regarding the states of the worker corresponds to the intangible social interaction responses. Such responses are often hard to directly detect and analyse. Nevertheless, thanks to the advances made in artificial intelligence emotional, cognitive and behavioural states of humans can now be inferred by physiological response monitoring with a reasonably good accuracy \cite{dinh2020stretchable}. Thus, such algorithms can be leveraged as an underlining technique of our approach to iteratively infer states of the worker while we tackle the problem of finding the best personalised action through an optimal policy given the states. Another important challenge in this RL approach is that of scallability. This is due to the fact that one must compute $Q(s,a)$ for every state-action pair in order to select the best action. This is computationally infeasible to compute when we have a larger state space. In recent RL works this issue has been addressed  by using a function approximator such as a neural network to approximate the action-value function. $Q(s,a;\theta)$ $\approx$ $Q^{*}(s,a)$ where $\theta$ is the function parameters(weights) of a neural network. Deep Q-learning is one of the most commonly used techniques to approximate optimal action-value functions using a deep neural network. This what Google's Deepmind used in ``AlphaGo"\cite{mnih2017methods}.  

Inspired by the practicality of such methods we define define our Q-function approximator using a neural network. This means in the forward pass of the network we use a loss function which tries to minimise the error of the Bellman equation. (\textit{i.e} determines how far $Q(s,a)$ is from the target $Q^{*}(s,a)$ given by:
\begin{equation}\label{eqn:Qv}
L_{i}(\theta_{i})  =  \mathbb{E}_{s,a\sim \rho(.) } \left [(y_{i} -Q(s,a;\theta_{i}))^{2} \right ]
\end{equation}
 where,
$y_{i} =  \mathbb{E}_{s^{'}\sim\varepsilon } \left [ r + \gamma\max_{a^{'}} Q(s^{'},a^{'};\theta_{i-1})|s,a \right ]$\\

The backward pass is then going to be a gradient update with respect to the Q-function parameters $\theta$. 

In summary, the personalisation framework can be divided in to three sequential layers to be implemented. The first layer has to do with identifying the main user of the personalisation service and main stakeholders as independent systems. This can be provided resorting the global formulation given in equation \ref{eqn:PuPSS} and translating it to the required context as done in equation \ref{eqn:PuCBx}. Once this is done in the second layer main objectives and the kinds of relations between the stakeholders will be identified as depicted in figure \ref{fig:fw}. This is useful to detect possible conflicts and interdependence among stakeholders. Subsequently the third layer formulates the problem of personalisation as an RL task. This provides an optimal operational policy for the personalising agent to actuate socially.

\section{Conclusion and Future work}
In this paper we proposed a  personalisation framework for Cyber-Physical-Social Systems(CPSS).
This is aimed at addressing the growing need to ensure a seamless human-machine interaction in the evolving smart environments which are conceived as CPSS. The framework was designed by first extending the systemic foundations of CPSS to characterise complex smart environments. The framework laid out in three different layers is believed to serve as a basis to design a more personalised and adaptable CPSS environments. In future work we plan to implement a method putting the mathematical formulations into practice. Especially by leveraging successful human state inference algorithms as an underlining technique. We believe that to ensure a seamless human-machine interaction finding optimal personalisation policies is a worthwhile endeavour. 
\bibliography{ifacconf}             
\end{document}